\documentclass[%
 reprint,showpacs,
 amsmath,amssymb, 
 aps,nofootinbib,superscriptaddress
]{revtex4-1}
 \usepackage{ulem}
    \usepackage{float}
\usepackage{graphicx, xcolor}
\usepackage{dcolumn}
\usepackage[colorlinks=true, linkcolor = red, citecolor = blue]{hyperref}
\usepackage{subcaption}

\usepackage{float}
\usepackage{xcolor}

\newcommand{\be}{\begin{equation}}  \newcommand{\ee}{\end{equation}}
\newcommand{\bea}{\begin{eqnarray}} \newcommand{\eea}{\end{eqnarray}}

\DeclareMathOperator{\erfc}{erfc}

\begin{document}
\newcommand{\lu}[1]{\textcolor{brown}{#1}}
\newcommand{\quita}[1]{\textcolor{orange}{#1}}
\newcommand{\JCH}[1]{\textcolor{blue}{#1}}
\newcommand{\DJM}[1]{\textcolor{red}{#1}}

\newcommand{\TDG}[1]{\textcolor{green}{#1}}
\title{Detecting the Stochastic Gravitational Wave Background from Primordial Black Holes in Slow-reheating Scenarios}

\author{Luis E. Padilla}
\email{l.padilla@qmul.ac.uk}
\affiliation{Astronomy Unit, Queen Mary University of London, Mile End Road, London, E1 4NS, UK.}

\author{Juan Carlos Hidalgo}
  \email{hidalgo@icf.unam.mx}
\affiliation{Instituto de Ciencias F\'isicas, Universidad Nacional Aut\'onoma de M\'exico, 62210, Cuernavaca, Morelos, M\'exico.}
  \author{Karim A. Malik}
  \email{ k.malik@qmul.ac.uk}
  \affiliation{Astronomy Unit, Queen Mary University of London, Mile End Road, London, E1 4NS, UK.}
  \author{David Mulryne}
  \email{d.mulryne@qmul.ac.uk}
  \affiliation{Astronomy Unit, Queen Mary University of London, Mile End Road, London, E1 4NS, UK.}

\date{\today}%

\begin{abstract}
After primordial inflation, the universe may have experienced a prolonged reheating epoch, potentially leading to a phase of matter domination supported by the oscillating inflaton field. During such an epoch, perturbations in the inflaton virialize upon reentering the cosmological horizon, forming inflaton structures. If the primordial overdensities are sufficiently large, these structures collapse to form primordial black holes (PBHs).  
To occur at a significant rate, this process requires an enhanced primordial power spectrum (PPS) at small scales.
The enhancement of the PPS, as well as the formation and tidal interaction of the primordial structures, will in turn source a stochastic gravitational wave background(SGWB)
that could be detected by current and/or future gravitational wave detectors. 
In this paper, we study the SGWB arising from these different sources 
during slow-reheating, focusing
on a PPS that satisfies the requirements necessary for the formation of PBHs with a mass of $M_{\rm PBH}\simeq 10^{21}$ and that constitute the entirety of dark matter in the universe. 
\end{abstract}

\maketitle

\section{Introduction}

Primordial black holes (PBHs) have been studied for more than half a
century \citep{zel1967hypothesis}. Despite the absence of definitive
proof for their existence, the recent advances in gravitational wave
astronomy have renewed interest in studying these objects.

The theory of Hawking radiation suggests that PBHs with masses $M_{\rm
  PBH}\lesssim 10^{15} \rm{g}$ should have evaporated by now
\citep{Hawking:1974rv}. On the other hand, PBHs with masses larger
than $10^{15} \rm{g}$ might constitute a portion of the dark matter
(DM) in the universe (see Ref.~\cite{Green:2020jor} for a recent review). Even if their contribution to the DM is not
substantial, their existence holds the potential to explain
observations like the precursors of supermassive black holes
\citep{PhysRevD.66.063505} or the presence of black holes within the
intermediate mass range \citep{sasaki2016primordial}. However, a mass
that proves to be particularly interesting to study is $M_{\rm PBH}\simeq
10^{21}$ g. This is because observations do not tightly constrain
these PBHs, so currently they could constitute the entirety of the DM
in the universe \citep{carr2010new,carr2020constraints}. Although PBHs
in the mass range $M_{\rm PBH}\simeq 10^{17}-10^{23}$ g could
potentially make up all the DM in the universe, we will focus in this
article on PBHs with a mass of $M_{\rm PBH}\simeq 10^{21}$ g for
simplicity as a representative example in the allowed range.

The cosmic microwave background (CMB) constrains the amplitude of the
primordial power spectrum (PPS) to be $\mathcal{P}_\zeta(k)\simeq 2.1
\times 10^{-9}$ at scales $k\sim 0.05~\rm{Mpc^{-1}}$
\citep{Planck:2018jri}. However, within the standard Big Bang
cosmology, in which a radiation-dominated (RD) universe is assumed right after the end of inflation, the formation of non-negligible abundances of PBHs requires amplitudes of the order $\mathcal{P}_\zeta(k) \sim 10^{-2}$. 
Consequently, numerous studies have explored inflationary
models capable of increasing the amplitude of primordial perturbations
at smaller scales (beyond those currently constrained by CMB results) to produce PBHs within
a desired mass range
\citep{Motohashi:2017kbs,PhysRevD.87.103506,Bugaev:2013fya,Erfani:2015rqv,Garcia-Bellido:2016dkw,Bezrukov:2017dyv}.

Various mechanisms have been proposed to achieve this amplification in
the PPS. These include postulating an inflection point
\citep{Garcia-Bellido:2017mdw} or a tiny bump
\citep{Mishra:2019pzq} in the inflaton potential. Additionally, in
multi-field models, amplification of perturbations is attainable
through isocurvature perturbations generated by non-inflaton field
dynamics (see e.g. 
Refs.~\citep{FUMAGALLI2023137921,PhysRevD.106.063535,Palma:2020ejf,Iacconi:2021ltm,Iacconi:2023slv,Kallosh:2022vha,Braglia:2022phb,fumagalli2023turning,braglia2020generating}. Although these mechanisms represent significant modifications to the simplest models of inflation, within this framework there may be no need to invoke physics beyond the standard model of particle physics to describe the nature of DM. 


In standard cosmological perturbation theory, scalar, vector, and
tensor perturbations decouple at first order. At higher-order, however, this is no
longer the case (see e.g.~Ref.~\cite{Malik:2008im}). For instance, a background of 
second-order tensor perturbations (gravitational waves) gets sourced by the square of
first-order scalar perturbations \cite{Ananda:2006af, Baumann:2007zm}. 
Despite their significant suppression at scales probed by the CMB 
due to the small amplitude of scalar perturbations on CMB scales, the amplification
of scalar perturbations necessary for PBH formation unavoidably generates a
substantial 
stochastic gravitational wave background (SGWB) on the scales associated with
PBH formation \cite{Domenech:2021ztg,Inomata:2023zup,Clesse:2018ogk,PhysRevD.95.043511,Nakama:2020kdc,PhysRevLett.126.051303,PhysRevLett.126.041303,PhysRevLett.126.041303}. Consequently, studying this
background also allows us to constrain models of the early universe
regarding PBH formation.


The physics of the early universe ranging from the end of inflation
to the onset of the Big Bang Nucleosynthesis (BBN), at a temperature
$T_{\rm BBN}\sim 10~\rm{MeV}$, is not very well constrained due to limited
observational probes of this
epoch. The cosmological standard model assumes that the universe quickly transitions to RD after inflation and remains in RD during this epoch. However, this scenario is not
inevitable, and there are several alternatives to this
description where the universe might have undergone different stages
of evolution \cite{Allahverdi:2020bys}. One of the most natural extensions 
is to consider the case of an early matter-dominated
(MD) epoch due to slow-reheating
\citep{carr1994black,LailaAlabidi_2013,PhysRevD.71.063507,PhysRevD.96.063504,Carrion:2021yeh}.

The energy scale at which the epoch of inflation is believed to have
occurred is $\lesssim 10^{14}~\rm{GeV}$ \citep{Planck:2018jri}. After inflation ends,  reheating must take place, {after which the universe becomes radiation dominated.} 
The efficiency of reheating
depends on the type of couplings the inflaton 
has with other particle species.
Reheating, therefore, may have been efficient
at energy scales much lower than those of inflation itself,
provided it was completed
before BBN. In such a slow-reheating 
scenario, the universe experiences a long MD epoch sourced by the oscillating inflaton field, during which there could have been different sources for GWs.

In this work, we consider the SGWB produced assuming an early MD epoch
generated by a slow-reheating phase.
During an MD epoch, the density
contrast grows proportionally to the scale factor, $\delta \propto
a$. As a consequence, 
{some scales get to reach a nonlinear regime}. This results in predictions for SGWB 
 from perturbation theory becoming 
unreliable at these scales, which is also the case for primordial structures and their interactions.
Consequently, the study of the
expected SGWB once an early MD epoch is included is typically
approached differently depending on the scales we are interested in. On
one hand, for those scales that do not reach the nonlinear regime,
calculating the SGWB at second order in perturbation theory should suffice to describe that part of the spectrum. On the other hand,
for the nonlinear scales, it is necessary to apply other techniques to correctly describe the SGWB.
Previous works have studied the SGWB from an early
MD and/or slow-reheating and produced intriguing findings. At the
perturbative level, it has been claimed that the details of the transition from the MD epoch to the RD epoch has a distinguishable impact. In the case of
a rapid transition (where the timescale is smaller than the Hubble
time), an increase in the energy density of GWs for modes that are
horizon-size at the transition time is expected
\citep{Inomata:2019ivs}. Conversely, a slow transition would lead to a
suppression of GWs for such scales \citep{Inomata:2019zqy}. For nonlinear scales, analytical estimates for the GWs
produced in an MD epoch were calculated in
Ref.~\citep{Karstenedamzik_2010}. These
estimates were tested and refined in Ref.~\citep{Eggemeier:2022gyo},
where a PPS produced by slow-roll inflation (without the features needed for PBH formation) was adopted and the SGWB
expected from slow-reheating phase studied.


Our work draws on and extends these earlier investigations by examining the SGWB
expected when PBHs form with a mass $M_{\rm PBH} \simeq
10^{21}~\rm{g}$. We choose this 
value since black holes formed with this mass during RD, constituting all of DM, produce a SGWB expected to peak within the maximum sensitivity range of LISA \citep{Bartolo:2018rku,Bartolo:2018evs}. 
Instead of formation during RD, however, we 
concentrate on the case in which such PBHs formed in a slow-reheating scenario. 
To do so we
will adopt the PBH formation criterion proposed in
\citep{Padilla:2021zgm} (see also
\citep{Hidalgo:2022yed,Padilla:2023lbv}), which has not previously been used in studies related to SGWB formation. {As shown in Ref.~\citep{Padilla:2021zgm}, this PBH formation criterion takes into account the quantum nature of the scalar field: In the hydrodynamic description of he post-inflationary universe, a quantum potential arises, which mimics a velocity dispersion at scales much larger than the de Broglie wavelength of the inflaton field. It is worth mentioning that this PBH formation criterion is the most restrictive compared to others proposed in the literature; therefore, similarly, it will be the most restrictive regarding the expected signal of the SGWB in this PBH formation scenario.}

{{This work also contributes to  the study of the effects on the SGWB at the transition to RD.} In particular, it has been shown that in the case of having a rapid transition from MD to RD, an increase in the amplitude of the SGWB is expected for those scales that become sub-horizon during MD. As for the case of a slow transition, the opposite is expected. In that resepect, our work shows a way to interpolate between these two previously reported results and allows studying more complicated scenarios in which the reheating of the universe could take place.}


The paper is organized as follows. In Sec.~\ref{Sec:II}, we review the
fundamentals of the slow-reheating scenario and PBH formation during
this epoch. In Sec.~\ref{Sec:III}, we present the type of PPS considered in
this work, while in Sec.~\ref{Sec:V} the expected spectrum of GWs is
presented for each of the examples studied. In Sec.~\ref{Sec:VI}, we
discuss the possibility of detection by earth- and space-based
experiments of the models presented in the preceding section. Finally, we concluding in 
Sec.~\ref{Sec:VII}.

\section{slow-reheating and the criterion for PBH formation}\label{Sec:II}

\subsection{The slow-reheating epoch}

After the inflationary epoch, the inflaton field undergoes a quick transition to a new phase of its evolution known as reheating, where it oscillates around the minimum of its potential, transferring energy to standard model particles. The simplest and most commonly used approximation for the inflaton dynamics around its minimum is to assume its potential can be approximated by a quadratic potential, i.e. 
\begin{equation}\nonumber
V(\varphi) = \frac{\mu^2}{2}\varphi^2\,,
\end{equation}
where $\mu$ is the mass of the inflaton field $\varphi$. In this case, the equation of state of the background universe can be approximated by $w = 0$, modelling the reheating phase as a MD universe in which the energy density of the cosmological background is expected to evolve according to
\begin{equation}\label{eq:background}
    \rho(a) \simeq \rho_{\rm end}\left(\frac{a_{\rm end}}{a}\right)^3\,.
\end{equation}
Hereafter the suffix ``{end}" refers to quantities evaluated at the end of inflation.

The early MD epoch of the universe continues as long as the Hubble parameter $H$ is larger than the decay rate $\Gamma$ of the inflaton field. When the condition $H\simeq \Gamma$ is met, the transfer of the energy associated with the inflaton field becomes efficient. The temperature at the time at which $\Gamma \simeq H$ is known as the reheating temperature and is given by $T_{\rm reh}\sim \sqrt{\Gamma M_{\rm Pl}}$, where $M_{\rm Pl}$ is the reduced Planck mass. Note, that if $\Gamma$ is small, then $T_{\rm reh}$ is also small.

During this epoch of the evolution of the universe, density perturbations associated with the smallest scales of the PPS reenter the cosmological horizon. For a given comoving wavenumber $k$ this occurs after
\begin{equation}
    N_{\rm HC}(k) = 2\ln\left(\frac{k_{\rm end}}{k}\right)
\end{equation}
efolds of expansion, where the suffix ``HC" denotes quantities evaluated at horizon crossing. After horizon crossing, the fluctuations are expected to grow as $\delta\sim a$ (where $\delta$ is the contrast density) until becoming nonlinear and forming primordial structures such as CDM-like structures or PBHs. The number of efolds necessary after the end of inflation to reach the nonlinear regime is given by \citep{Hidalgo:2022yed}
\begin{equation}\label{eq:nl}
    N_{\rm NL}(k) = N_{\rm HC}(k)+\ln[1.39\delta_{\rm HC}^{-1}(k)]\,.
\end{equation}
A requirement for the primordial structures to form is therefore
$N_{\rm NL}(k)\leq N_{\rm reh}$, where
\begin{equation}
    N_{\rm reh} \simeq \frac{1}{3}\ln\left[ \frac{A_s}{2}\left(\frac{90}{g_{\rm reh}}\right)\frac{M_{\rm Pl}^4 r}{T_{\rm reh}^4}\right]\,
\end{equation}
is the number of e-folds the reheating epoch lasted. 
To obtain the above expression we have considered that $N_{\rm reh} = (2/3)\ln(H_{\rm end}/H_{\rm reh})$, where $H_{\rm reh}^2 = \rho_{\rm reh}/(3M_{\rm Pl}^2)$, $\rho_{\rm reh}$ is the energy density at the reheating time ($\rho_{\rm reh} = \pi^2(g_{\rm reh}/30)T_{\rm reh}^4$, where $g_{\rm reh}$  is the effective number of relativistic species and $T_{\rm reh}$ is the reheating temperature), and $H_{\rm end}\simeq \sqrt{\pi^2 M_{\rm Pl}^2 A_s r/2}$, with $A_s\simeq 2.0989\times 10^{-9}$ being the amplitude of the scalar perturbations measured at the CMB scales and $r<0.068$ the tensor to scalar ratio \citep{Planck:2018jri}. In what follows, we assume the upper limit for $r$. {However, in the last part of this article we elaborate more on the consequences of considering a smaller value of $r$, as it is forecasted in future surveys.} 

\subsection{Abundance of primordial black hole formation during slow-reheating}\label{Sec:IV}

In the simplest and most commonly studied scenario of PBH formation, density fluctuations with a large amplitude reenter the cosmological horizon in post-inflationary epochs and collapse to form PBHs (we focus on PBHs forming at horizon entry, for sub-horizon PBHs see Refs.~\citep{Zaballa:2006kh, Lyth:2005ze, Torres-Lomas:2014bua}). The mass of such PBHs should be close to the mass of the cosmological horizon evaluated at the time of horizon crossing, $M_{\rm PBH}(k) = \gamma M_{\rm HC}(k)$, where $\gamma$ is a constant that encodes the efficiency of the collapse\footnote{In the standard RD case $\gamma \simeq 0.2$ \citep{PhysRevD.81.104019,Carr_2021}. The specific value of $\gamma$ in a collapse scenario related to slow-reheating is not well known, though partial progress has been reported in Refs.~\citep{deJong:2021bbo,Padilla:2021uof}. Due to this uncertainty, we take the conservative value of $\gamma = 1$, considering that the results obtained will not be strongly influenced by the particular value of $\gamma$.}. The PBH mass can thus be expressed as
\begin{equation}\label{mass}
\frac{M_{\rm PBH}(k)}{10^{21}\rm{g}} \simeq \gamma\left[\left(\frac{1.33\times 10^{-7}~\rm{GeV}}{H_{\rm ref}}\right)\left(\frac{k_{\rm ref}}{k}\right)^{3}\right],
\end{equation}
where $H_{\rm ref}$ ($k_{\rm ref}$) is a Hubble ($k$) reference scale evaluated at a time within the slow-reheating epoch (it can be, for example, the time at the RD transition). From the above expression, it is also evident that for PBHs with a mass of $M_{\rm PBH} \simeq 10^{21}~\rm{g}$ to form within the slow-reheating scenario, it is necessary for the Hubble parameter at the transition/reheating time to be smaller than $H_{\rm{reh}} = 1.33\times 10^{-7}~\rm{GeV}$. This particular value corresponds to a maximum reheating temperature of $T_{\rm reh} \simeq 3\times 10^5~\rm{GeV}$\footnote{This maximum temperature assumes PBHs form immediately after horizon crossing. However, as can be seen from Eq.~\eqref{eq:nl} we need perturbations to grow and reach the nonlinear regime for them to collapse to form PBHs. This implies the maximum temperature for reheating should be a slightly smaller.}.   

We compute the abundance of PBH using the Press-Schechter formalism \citep{Press:1973iz}. The fraction of the total energy content in PBHs is then given by
\begin{equation}\label{eq:abundance}
    \beta(M) =-2M\frac{\partial R}{\partial M}\frac{\partial P[\delta>\delta_{c}]}{\partial R}\,, 
\end{equation}
where 
\begin{equation}
    P[\delta>\delta_{c}] = \frac{1}{2}\erfc\left(\frac{\delta_{ c}}{\sqrt{2}\sigma(R)}\right) \,,
\end{equation}
$\delta_c$ is the collapse threshold for PBHs to form, and $\sigma^2(R)$ is the standard deviation of the  density contrast $\delta$ evaluated at the horizon crossing time,
\begin{equation}
    \sigma(R)^2 = \left[\frac{2(1+\omega)}{5+3\omega}\right]^2\int_0^\infty W^2(\tilde k R)\left(\frac{\tilde k}{aH}\right)^4\mathcal{P}_{\mathcal{R}} (\tilde k,t_{\rm HC})d\ln\tilde k\,.
\end{equation}
In the above equation $W(kR) = \exp(-k^2R^2/2)$ is the Fourier transform of the window function used to smooth the density contrast over
a scale $R = 1/k$ and $\omega$ is the equation of state of the background universe, i.e.~during slow-reheating $\omega = 0$.

As discussed in Ref.~\citep{Padilla:2024iyr} the threshold value $\delta_c$ in slow-reheating depends on different particularities that the initial perturbation may have, which we will describe in what follows. First of all, in the context where the reheating of the universe is effectively assumed as a pressureless fluid, one could simply extrapolate the perfect fluid criterion $\lim_{\omega\rightarrow 0}\delta_c \ \rightarrow \ \omega$ \citep{Harada:2013epa} and obtain a PBH formation criterion during slow-reheating. However, even in this simple description, it has been shown that such a criterion would overestimate the expected abundance of PBHs, as effects like the non-sphericity of the initial perturbation \citep{Harada:2016mhb} or its angular momentum \citep{Harada:2017fjm} must play a significant role in preventing the gravitational collapse into PBHs. On the other hand, given that in reheating, the component that is generating the early MD phase is a scalar field, it is generally not possible to disregard the quantum nature of the field. Instead, it is necessary to understand how this property plays a role in the generation of PBHs. In particular, the pressure caused by Heisenberg's uncertainty principle (which would be a pressure that prevents collapse) can be approximated as an effective velocity dispersion that the collapsing system would have. This effective velocity dispersion can be so efficient in preventing collapse that it could even be more restrictive than the previously mentioned criteria (for $\sigma\leq 0.04$). With this in mind, the criterion for PBH formation that we will use in the remainder of this work will be that provided by the effective velocity dispersion, in which the formation of PBHs is assumed to occur whenever an initial overdensity exceeds the threshold value \citep{Padilla:2021zgm}:
\begin{equation}\label{eq:thr}
    \delta_c \simeq 0.238\,.
\end{equation}
Note, that despite the above threshold being lower than the value necessary to form PBHs during the standard RD era ($\delta_c^{\rm rad} = 0.41-0.66$, \citep{Green:2004wb,PhysRevD.100.123524,PhysRevD.101.044022,Escriva:2020tak}), the difference is small, hence to form a significant number of PBHs, it is still necessary for the PPS amplitude to increase to roughly $\mathcal{P}_\zeta(k)\sim 10^{-2}$.

\section{Parameterizing the primordial power spectrum}\label{Sec:III}

In this article, we shall work with a log-normal peak power spectrum,
\begin{equation}\label{eq:spectrums}
    \mathcal{P}_\mathcal{R}(k) = A_s\left(\frac{k}{k_*}\right)^{n_s-1}+B_p \exp\left[-\frac{\ln^2(k/k_p)}{2\sigma_p^2}\right].
\end{equation}
The first term on the r.h.s. of the above equation corresponds to the usual slow-rolling power spectrum,\footnote{The free parameters of the slow-rolling power spectrum are the amplitude of scalar perturbations, $A_s$, and the spectral index, $n_s$, typically constrained around the pivot scale $k_* =0.05~\rm{Mpc}$.} typically constrained by CMB physics. For definiteness, in this article, we use the Planck preferred values \citep{Planck:2018jri} $A_s = 2.0989\times 10^{-9}$ and $n_s = 0.9649$. On the other hand, the second term (the log-normal peak) is the parameterized power spectrum corresponding to a feature at the smallest scales, necessary for the formation of PBHs. The parameters $B_p$, $\sigma_p$, and $k_p$ controls the amplitude, width, and position of the peak in the PPS.

The log-normal peak power spectrum typically emerges in some models of single field inflation with an inflection point \citep{germani2017primordial} or in some versions of smooth-waterfall hybrid inflation \citep{clesse2015massive}. Additionally, in the limit $\sigma_p\rightarrow 0$ such a parametrization reduces to a sharp peak power spectrum, expected in some inflationary models such as hybrid inflation \citep{Garcia-Bellido:1996mdl,Lyth:2012yp} or axion inflation with couplings to gauge fields \citep{linde2013gauge,Bugaev:2013fya}. For simplicity, in this article we will limit ourselves only to studying the case $\sigma_p = 0.5${, which is a choice consistent with the requirement that the PPS can not easily grow with scale faster than $k^4$ in single-field models inflationary models \citep{Byrnes:2018txb,Carrilho:2019oqg}}. This leaves us with two free parameters in the model, namely, $B_p$ and $k_p$, which will be adjusted for different examples.

\begin{figure}
\centering
    \includegraphics[width=3.in]{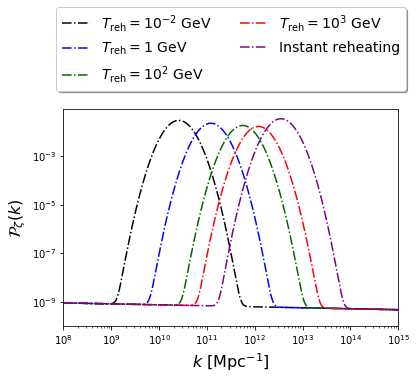}
    \includegraphics[width=3.in]{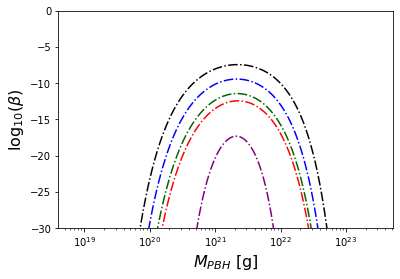}
    \caption{\footnotesize{Upper Panel: Primordial power spectrum required to form PBHs with a mass of $M_{\rm PBH}\simeq 10^{21}~\rm{g}$, which  constitute the totality of the DM in the universe; different reheating temperatures are considered. Lower Panel: Abundance of PBHs at formation time for each case.}}
    \label{fig:pps_and_beta}
\end{figure}

We plot in Fig.~\ref{fig:pps_and_beta} the PPS \eqref{eq:spectrums} and the abundance of PBHs at formation time necessary for PBHs with a mass of $M_{\rm PBH}\simeq 10^{21}~\rm{g}$ to constitute the entirety of the DM in the universe. With such condition, we sample a range of reheating temperatures. Note that, excluding the purple curve, the reheating temperature in each scenario meets the condition $T_{\rm reh}< 3\times 10^{5}~\rm{GeV}$, {which proves the viability of PBHs of interest to be formed during an early slow-reheating epoch {(see paragraph just after Eq.~\eqref{mass})}}. Thus, in order to account for PBHs in these scenarios, we must employ threshold values given in Eq.~\eqref{eq:thr}. To contrast our results with the standard scenario, we consider the case of an instant reheating, represented with a purple line. In that case, PBHs should form during RD and then we used the threshold value $\delta_c = 0.41$ as an example. To accurately determine the abundance of PBHs at their formation, such that they could presently constitute the entire mass of the DM in the universe, we use the freely available \texttt{betaPBH} code \citep{Gomez-Aguilar:2023bej}, designed to calculate constraints on PBH abundance in scenarios involving standard and non-standard cosmologies such as an early MD epoch.

It is interesting to observe from the figure that, despite having abundances of PBHs that vary greatly for each reheating temperature, the primordial power spectrum is not strongly modified. Instead, having a different reheating temperature shifts the position of the peak in the PPS. This shift will thus be crucial to study when connecting these PPSs with the generation of the SGWB for each studied example.


\section{Stochastic gravitational wave background}\label{Sec:V}

\begin{figure}
    \centering
    \includegraphics[width=3.5in]{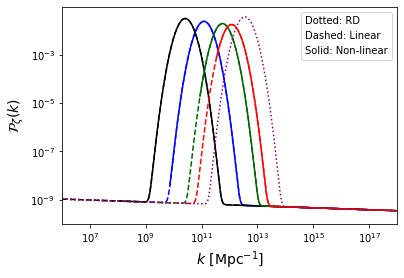}
    \caption{\footnotesize{PPS for all the models \lu{shown in Fig.~\ref{fig:pps_and_beta}}.
    We divided the spectrum into those scales that re-enter the cosmological horizon and reach the nonlinear regime (solid curve), those that remain within the perturbative regime (dashed curve), and those scales that continue to be super-horizon throughout the reheating period (dotted).}}
    \label{fig:hc}
\end{figure}

As previously argued, it is expected that during slow-reheating some scales may have grown sufficiently to form primordial structures such as CDM-like structures or PBHs. The condition for this to happen is that $N_{\rm HC}(k)\leq N_{\rm reh}$ (for a scale to re-enter the cosmological horizon during slow-reheating) and $N_{\rm NL}(k)\leq N_{\rm reh}$ (for the scale to form structure). As an example, in Fig.~\ref{fig:hc}, we have plotted again all of the PPS shown in Fig.~\ref{fig:pps_and_beta} but this time
we indicate with a solid curve the scales that would reach the nonlinear regime and we expect to form primordial structures (where we have used the relation $\delta_{\rm HC}(k)\simeq (2/5)\mathcal{P}_\mathcal{R}(k)^{1/2}$), with a dashed line for those scales that re-enter the cosmological horizon but do not reach the nonlinear regime, and with a dotted line for those scales that remain super-horizon throughout reheating. In what follows, we will calculate the expected SGWB for each of the examples shown in Fig.~\ref{fig:pps_and_beta}, studying differently the spectrum for the nonlinear scales and those that do not reach this regime. It is necessary to clarify also that for PBHs to be formed from the peak of the PPS, said peak must be in the region of scales that reach the non-linear regime, so the majority contribution of the expected SGWB due to the peak will be that which is generated due to the non-linear dynamics of the formation and interaction of the primordial structures.

\subsection{Second order SGWB contributions}\label{sec:second_o}

As mentioned before, the generation of GWs in the perturbative regime is closely linked to the duration of the transition from the early MD phase (caused by the inflaton field in our scenario) to the RD epoch. In this subsection, we will start by computing the most common scenario of an instantaneous transition from MD to RD, closely following Ref.~\citep{Inomata:2019ivs}. Then we will expand our analysis to include more gradual transitions. As a bonus, we will provide a general framework for interpolating between the gradual transition discussed in Ref.~\citep{Inomata:2019zqy} and the rapid transition, Ref.~\citep{Inomata:2019ivs}. 

\subsubsection{Instant transition scenario}

We start by reviewing the general formulas needed to calculate the SGWB in the perturbative regime. Assuming an instant transition from the slow-reheating to the RD epoch at the conformal time $\eta = \eta_{\rm reh}$ ($\eta = \int dt/a$), we can write the time evolution of the scale factor during and after the reheating of the universe as
\begin{equation}
    \frac{a(\eta)}{a(\eta_{\rm reh})} = \begin{cases}
        \left(\frac{\eta}{\eta_{\rm reh}}\right)^2 & \eta\leq \eta_{\rm reh}\\
        2\frac{\eta}{\eta_{\rm reh}}-1 & \eta> \eta_{\rm reh}.
    \end{cases}
\end{equation}

At second order in perturbation theory, the energy density parameter of GWs can be defined as \citep{Kohri:2018awv}
\begin{equation}
    \Omega_{\rm GW}(\eta, k) = \frac{1}{24}\left(\frac{k}{a(\eta)H(\eta)}\right)^2\overline{\mathcal{P}_h(\eta,k)}\,,
\end{equation}
where $\overline{\mathcal{P}_h(\eta,k)}$ is the time averaged power spectrum of GWs,
\begin{eqnarray}
    \overline{\mathcal{P}_h(\eta,k)} = &&4 \int_0^{\infty}dv \int_{|1-v|}^{1+v}du \left(\frac{4v^2-(1+v^2-u^2)^2}{4vu}\right)^2\nonumber \\
    &&\times \overline{I^2(u,v,k,x,x_{\rm reh})}\mathcal{P}_\zeta(uk)\mathcal{P}_\zeta(vk)\,, \label{eq:P_h}
\end{eqnarray}
where $x =k\eta$ and $x_{\rm reh} = k\eta_{\rm reh}$. The integral ${I}$ includes the time-dependence of GWs and can be written as
\begin{equation}
    I(u,v,x,x_{\rm reh}) = I_{\rm reh}(u,v,x,x_{\rm reh})+I_{\rm RD}(u,v,x,x_{\rm reh})\label{eq:I},
\end{equation}
where
\begin{subequations}
\begin{eqnarray}
    I_{\rm reh}(u,v,x,x_{\rm reh})=&& \int_0^{x_{\rm reh}}d\bar x\left(\frac{1}{2(x/x_{\rm reh})}-1\right)\left(\frac{\bar x}{x_{\rm reh}}\right)^2\nonumber \\
    &\times& kG_k^{\rm reh}(\eta,\bar\eta)f(u,v,\bar x,x_{\rm reh})\nonumber
\end{eqnarray}
and
\begin{eqnarray}
    I_{\rm RD}(u,v,x,x_{\rm reh}) =&&\int_{x_{\rm reh}}^xd\bar x\left(\frac{2(\bar x/x_{\rm reh}-1)}{2(x/x_{\rm reh}-1)}\right)\nonumber \\
    &\times& kG_k^{\rm RD}(\eta,\bar\eta)f(u,v,\bar x,x_{\rm reh}).\nonumber
\end{eqnarray}
\end{subequations}
In the above expressions, $I_{\rm reh}$ and $I_{\rm RD}$ represent the contribution of GWs induced during the slow-reheating and the RD epochs, respectively, $G_k^{\rm reh}$ and $G_k^{\rm RD}$ are two Green functions valid at the slow-reheating and RD scenarios, and $f$ is a source function defined as
\begin{eqnarray}
    f(u,v,\bar x,x_{\rm reh})=&&\frac{3}{25(1+\omega)}[2(5+3\omega)\Phi(u\bar x)\Phi(v\bar x)\nonumber \\
    &+& 4\mathcal{H}^{-1}(\Phi^{'}(u\bar x)\Phi(v\bar x)+\Phi(u\bar x)\Phi^{'}(v\bar x))\nonumber \\
    &+&4\mathcal{H}^{-2}\Phi^{'}(u\bar x)\Phi^{'}(v\bar x)].\label{eq:f}
\end{eqnarray}
In the above $\mathcal{H}$ is the conformal Hubble parameter and $\Phi(x)$\footnote{Note, that we used $\Phi(x)$ as a short notation for $\Phi(x,x_{\rm reh})$.} is the gravitational potential, which in the instant reheating scenario is given by
\begin{equation}\label{eq:pot_inst}
    \Phi(x)/\Phi_0 =\begin{cases}
        1 & x\leq x_{\rm reh}\\
        A(x_{\rm reh})\mathcal{J}(x)+B(x_{\rm reh})\mathcal{Y}(x) & x>x_{\rm reh}\,. 
    \end{cases}
\end{equation}
The coefficients $A(x_{\rm reh})$ and $B(x_{\rm reh})$ are determined by imposing continuity of $\Phi(x)$ and $\Phi^{'}(x)$ at $\eta_{\rm reh}$, $\mathcal{J}(x)$ and $\mathcal{Y}(x)$ are two functions defined in terms of spherical Bessel functions of the first and second kind (see Ref.~\citep{Inomata:2019ivs}), and $\Phi_0$ is the value of the gravitational potential at the beginning of MD. 

With these definitions, we can approximate $\overline{I^2}$ in Eq.~\eqref{eq:P_h} as
\begin{equation}\label{eq:I}
    \overline{I^2(u,v,k,x,x_{\rm reh})}\simeq \overline{I_{\rm reh}^2(u,v,k,x,x_{\rm reh})}+\overline{I_{\rm RD}^2(u,v,k,x,x_{\rm{reh})}}\,.
\end{equation}
The functional form of $\overline{I^2_{\rm reh}}$ can be found in Ref.~\citep{PhysRevD.97.123532} whereas the approximations for $\overline{I^2_{\rm RD}}$ were given in Appendix B of Ref.~\citep{Inomata:2019ivs}.

We have calculated numerically the SGWB expected in each of the examples shown in Fig.~\ref{fig:pps_and_beta}. For such a calculation, we have used the PPS 
\begin{eqnarray}
        \mathcal{P}_\mathcal{R}(k) = &&\left(A_s\left(\frac{k}{k_*}\right)^{n_s-1}+\left.B_p \exp\left[-\frac{\ln^2(k/k_p)}{2\sigma_p^2}\right]\right)\right.\nonumber \\
        &\times&\Theta(k_{\rm max}-k)\label{eq:pps2}
\end{eqnarray} 
in Eq.~\eqref{eq:P_h}, where the Heaviside step function $\Theta(k_{\rm max}-k)$ was included to give a cutoff at $k_{\rm max}$, where, the maximum wavenumber expected to become non-linear during the slow-reheating epoch, computed by solving the equation $N_{\rm NL}(k_{\rm max}) = N_{\rm reh}$. 

\begin{figure}
    \centering
    \includegraphics[width=3.5in]{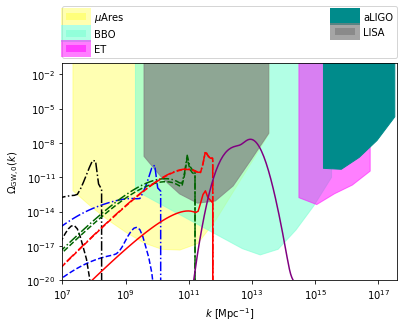}
    \caption{\footnotesize{SGWB at second-order in perturbations theory expected in a slow-reheating scenario. Each color follows the description of Fig.~\ref{fig:pps_and_beta}. The solid, dashed, and dot-dashed lines correspond to the values $\beta = 0.1$, $10^{-3}$, and the instant-transition scenario, respectively (see main text). To contrast our scenarios of slow-reheating, we have also included the purple curve (of instant reheating) which was calculated with SIGWfast \citep{Witkowski:2022mtg}. For reference we plot the current and future sensitivity curves of space- and earth-based experiments (see Ref.~\citep{Campeti_2021}).}}
    \label{fig:SGWB_SO}
\end{figure}

Our results on GWs energy fraction are depicted in Fig.~\ref{fig:SGWB_SO} using dot-dashed lines where, to obtain the current energy density parameter of GWs, $\Omega_{\rm GW,0}(k)$, we have taken the relation
\begin{equation}
    \Omega_{\rm GW,0}(k) = 0.39\left(\frac{g_c}{106.75}\right)^{-1/3}\Omega_{r,0}\Omega_{\rm GW}(\eta_c,k).
\end{equation}
In the above expression $\Omega_{r,0}$ is the current energy density parameter of radiation, and $g_c$ and  $\Omega_{\rm GW}(\eta_c,k)$ are the values of the effective relativistic degrees of freedom and the energy density of GWs at the time at which the gravitational potential has sufficiently decayed (during RD) and $\Omega_{\rm GW}$ becomes constant. The figure also shows sensitivity curves for the energy density of GWs from experiments such as LISA, BBO, advanced LIGO, $\mu$Ares, and the Einstein Telescope (ET). As illustrated, our findings suggest potential detection signals across various experiments. This is primarily attributed to the \textit{Poltergeist mechanism}, which arises from an instantaneous transition between the slow-reheating and the RD epoch, and is associated to the term $\bar I^2_{\rm RD}$ in Eq.~\eqref{eq:P_h}. As discussed in Ref.~\citep{Inomata:2019ivs}, the amplification of induced GWs at small scales ($k\gg 1/\eta_{\rm reh}$, where scales are subhorizon before the slow-reheating to RD transition) is due to the time independence of the gravitational potential until $\eta = \eta_{\rm reh}$ (see Eq.~\eqref{eq:pot_inst}). Subsequently, {after reheating?}, the potential begins to oscillate with a timescale $\sim 1/k$, significantly shorter than its decay timescale $\eta_{\rm reh}$. This rapid oscillation, combined with the substantial amplitude of the gravitational potential at the transition time, contributes to the amplification of GWs. 

\subsubsection{The case of a gradual transition}\label{sec:gradual}

The instant transition scenario is a special case.
For example, the inflaton field can decay into radiation also through a time-dependent decay rate $\Gamma(\eta)$. In this case, the main difference between a rapid transition and a gradual one from early-reheating to RD is related to how the gravitational potential evolves during the transition (after all, the gravitational potential is the main source of second-order GWs). As we saw earlier, in a rapid transition, the gravitational potential is expected to start oscillating abruptly with rapid oscillations and an amplitude $\Phi\sim \Phi_0$ immediately after reaching the time $\eta = \eta_{\rm reh}$ (see Eq.~\eqref{eq:pot_inst}). However, in a more gradual transition the gravitational potential decreases on sub-horizon scales (following the evolution of matter perturbations) before starting its phase of rapid oscillations (once matter perturbations have decayed enough for them to be negligible). This decrease affects the spectrum of SGWB expected in these scenarios. In this section, we will provide a general description of how to estimate the SGWB for scenarios of a gradual transition between the early slow-reheating and the late RD universe, and then apply this formalism to the different examples studied in this work.

According to Ref.~\citep{Inomata:2019zqy}, we can approximate the time evolution of the gravitational potential during the transition from the early reheating epoch to RD as
\begin{eqnarray}
    \Phi(\eta)/\Phi_0 &\simeq& \exp\left(-\int_{\eta_i}^{\eta}d\bar\eta a(\bar \eta)\Gamma(\bar \eta)\right),\nonumber \\
    &\simeq& \exp(-\alpha(\eta)\bar\Gamma t),\nonumber \\
    &\simeq& \exp\left(-2\alpha(\eta) t/(3t_{\rm reh})\right). \label{eq:phi_eta}
\end{eqnarray}
where we have taken $\bar \Gamma = 2/(3t_{\rm reh}),$ $\bar\Gamma$ is a constant decay rate (i.e.~$\Gamma = \bar\Gamma f(\eta)$), $\alpha$ is a function that encodes the functional form of the decay rate,
\begin{equation}
    \alpha(\eta) = 1+\frac{1}{ t(\eta) f(\eta)}\int_{\eta_i}^{\eta}d\bar \eta a(\bar \eta) t(\bar \eta)\frac{df(\bar\eta)}{dt}
\end{equation}
and $t$ is the cosmic time,
\begin{equation}
    t = \int_{\eta_i}^{\eta }d\bar\eta a(\bar \eta).
\end{equation}
Note, that in the limit $\alpha(\eta) t/t_{\rm reh} \rightarrow 0$, the gravitational potential does not decrease during reheating, so we take this limit as the limit of {instantaneous transition?}. On the other hand, when $\alpha = 1$, we obtain $\Gamma = \bar\Gamma$, representing a constant decay rate, typically predicted in perturbative reheating scenarios by an exactly quadratic potential (see for example Ref.~\citep{Garcia:2020wiy}). Since the nature of reheating is not well understood, we will not consider a particular form for $\alpha$ for the meantime.

We will study next how this decay of the gravitational potential affects the expected SGWB in a more gradual reheating scenario. To this end, we follow a similar procedure as in Ref.~\citep{Inomata:2020lmk}, where the authors were interested in calculating the SGWB expected during reheating by PBH evaporation. Given that in the instant transition scenario the main contribution of GWs is produced by the GWs generated during RD (the term $\overline{I^2_{\rm RD}}$ in Eq.~\eqref{eq:I}), we can approximate the expected GWs during a gradual transition by focusing solely on $I_{\rm RD}$ (i.e.~we assume that the GWs generated during the slow-reheating epoch are negligible compared to those generated during the RD epoch). We also assume the gravitational potential follows Eq.~\eqref{eq:phi_eta} until the decoupling time $\eta_d$ at which the gravitational potential starts its phase of fast oscillations. Then, after $\eta_d$ we can approximate the evolution of the gravitational potential by
\begin{equation}\label{eq:ph_}
    \Phi(x>x_d)\simeq T(\alpha,k)\Phi_{\rm inst}(x>x_d)\,,
\end{equation}
where $T(\alpha,k)$ is a transfer function given by
\begin{equation}\label{eq:tf}
    T(\alpha,k) = \exp(-2\alpha(\eta_d)t_d/(3t_{\rm reh}))\,,
\end{equation}
and where the decoupling time and the $\alpha$ parameter generally depend on the scale $k$. In Eq.~\eqref{eq:ph_} $\Phi_{\rm inst}$ is the value of the gravitational potential valid in the instant transition scenario, once we replace $\eta_{\rm reh}$ by $\eta_{d}$.

Substituting Eq.~\eqref{eq:ph_} into Eq.~\eqref{eq:f} and performing the integral in Eq.~\eqref{eq:I}, we arrive at 
\begin{eqnarray}
    \overline{I^2(u,v,k,\eta,\eta_R)}&\simeq& \overline{I^2_{\rm RD}(u,v,k,x,x_d)}\nonumber \\
    &\simeq& T(\alpha,uk)T(\alpha,vk)\nonumber \\
    &&\times\overline{I^2_{\rm RD,inst}(u,v,k,x,x_{ d})}\,,
    \label{eq:I_rd}
\end{eqnarray}
where $\overline{I^2_{\rm RD,inst}}$ accounts for $I$ for the instant transition scenario. 

The idea behind the above approximation is that the main contribution to the enhancement of GWs in the instant transition is the rapid oscillations of the gravitational potential at large amplitude. In this rapid scenario, what we calculate, essentially, is the amount of the GWs generated by these oscillations. However, by employing the aforementioned approximation, we also account for the contribution of the oscillatory component of the gravitational potential while considering the suppression resulting from the decay of the gravitational potential during the gradual transition from early reheating to RD.

Note that above we incorporated a global exponential suppression factor (dependent on the scale $k$ and the functional form of the parameter $\alpha$), which enables us to also calculate the SGWB expected in scenarios involving more gradual transitions. What we need to do next is determine the transfer function defined in Eq.~\eqref{eq:tf}.

We note, that Eq.~\eqref{eq:phi_eta} is valid as long as the condition $|\ddot \Phi/\Phi|\ll k^2/(3a^2)$ is fulfilled, where a period indicates a derivative with respect to cosmic time. Therefore, a good indicator for the validity of Eq.~\eqref{eq:phi_eta} is when the previous condition is no longer met, which we take to be when $|\ddot \Phi/\Phi|_{t = t_d}\simeq k^2/(3a(\eta_d)^2)$. In reality, we expect the approximation to cease being valid even before this equality is reached, so the results shown here regarding the SGWB should be taken as pessimistic predictions of the spectrum (however, see the discussion in Appendix~\ref{ap:1b} about Fig.~\ref{fig:inst_grad}). Since at this point we do not expect to assume a particular form of $\alpha$ (except for the case $\alpha = 1$ that reduces to a constant decay rate), we can instead consider the following simplification. First, in the case of $\alpha = 1$ we have
\begin{equation}
    \frac{t_d(k)}{t_{\rm reh}} \simeq \frac{1}{4}\left( \frac{x_{\rm reh}^2}{4}+1\right)\,.
\end{equation}
Then, we take as an ansatz the following relation 
\begin{equation}\label{eq:beta_par}
     \alpha(\eta_d)\frac{t_d(k)}{t_{\rm reh}} \simeq \beta\frac{1}{4}\left( \frac{x_{\rm reh}^2}{4}+1\right),
\end{equation}
i.e.~in the above relation, the new $\beta$ parameter encodes the functional form of $\alpha$ and is also  related to the functional form of $\Gamma$. In what follows, we will assume $\beta$ is a constant quantity in the range $[0,1]$, from which we can identify the case $\beta = 1$ as the limit of a constant decay rate (which was referred to in Ref.~\citep{Inomata:2019zqy} as the regime of gradual transition) and $\beta = 0$ as the limit of instant reheating. Despite the simplicity of our assumption for $\beta$, by assuming different values for $\beta$ we are taking into account that the gravitational potential began its regime of rapid oscillations at higher values, where the case $\beta = 1$ should be taken as the limiting case where the gravitational potential decayed as much as possible before starting its regime of rapid oscillations, while the case $\beta = 0$ assumes that the gravitational potential did not decay at all. 

To ensure that our approximations for $\beta$ are sufficient to correctly describe a more gradual transition from early reheating to RD, we have also added appendix~\ref{ap:1b}, where we calculate the SGWB for a slow-roll PPS and compare it with the results of Ref.~\citep{Pearce:2023kxp}. As we describe in the appendix, although we both start from different approximations to deal with transitions that interpolate between the gradual and the instantaneous transition, we both reproduce a similar SGWB, with differences only in the part of the spectrum that cannot be detected by the various experiments proposed to detect GWs.


Using the above equation into Eq.~\eqref{eq:tf} and Eq.~\eqref{eq:I_rd}, we calculate numerically the SGWB expected for different realizations of $\beta$ and for the same PPS shown in Eq.~\eqref{eq:pps2}. In particular, we plotted in Fig.~\ref{fig:SGWB_SO} the cases of $\beta = 0.1$ (solid) and $10^{-3}$ (dashed). In the particular case of $\beta = 1$, which corresponds to the case of a constant decay rate, the SGWB we found is very small (not detectable by any experiment), so we decided not to include it in our figure.

Note, that for the reheating temperatures $T_{\rm reh} = 10^2~\rm{GeV}$ (green) and $T_{\rm reh} = 10^3~\rm{GeV}$ (red), we found that for $\beta = 10^{-3}$ the plots are very close to the instant transition result, which is not the case for  $T_{\rm reh} = 10^{-2}~\rm{GeV}$ and $T_{\rm reh} =1~\rm{GeV}$. Furthermore, for $T_{\rm reh} = 10^3~\rm{GeV}$ with $\beta = 0.1$, we still have an increase in the spectrum that can be detectable for various experiments, which implies that having a reheating slightly faster than that given by a constant decay rate would be enough to produce a SGWB signal large enough for it to be detectable. This is a consequence of part of the peak in the PPS being within the subhorizon scales that do not manage to form structure ({see Fig.~\ref{fig:hc}}), so the increase in the SGWB spectrum for these scales is not only a consequence of the Poltergeist mechanism (as described in the previous subsection) but also due to the amplification of the PPS in that scale region.

\subsection{Non-linear dynamics contribution}

In this section, we estimate the expected SGWB for scales that become nonlinear during the epoch of slow-reheating. For these scales, we anticipate the main contribution to the GWs would come from the formation and tidal interaction of the primordial structures expected during this epoch (PBHs and CDM-like structures). 

To accurately calculate the SGWB in this scenario, we closely follow Ref.~\citep{Eggemeier:2022gyo} (see also Ref.~\citep{KarstenJedamzik_2010} for earlier work), where the radiated energy density of GWs generated by the fragmentation of the inflaton field into a primordial structure is estimated by
\begin{equation}
    h\simeq \frac{G}{2}\left(\ddot I_{ij}-\frac{1}{3}\ddot I_kk \delta_{ij}\right)\frac{n_in_j}{|\textbf{x} |}\,,
\end{equation}
where $h$ is the amplitude of the GW, $\textbf{n}$ is the radial unit vector from the origin located in the center of the source to the point $\textbf{x}$, and $I_{ij}$ is the quadrupole tensor. Assuming that a primordial structure with a mass $M_p$, radius $R_p$, and virial velocity $v_p = (GM_p/R_p)^{1/2}$ is formed, an order of magnitude estimate for $\ddot I_{ij}$ is
\begin{equation}
    \ddot I_{ij} = \frac{d^2}{dt^2}\int \rho x^2d^3x\sim 2M_pv_p^2\,.
\end{equation}
The above equation implies that $h\sim GM_pv_p n_i n_j/|\textbf{x}|$.

We can also estimate the energy flux of the GW as
\begin{equation}
    L_{\rm GW}^{\rm coll}\simeq \frac{|\textbf{x}|\dot h^2}{G}\sim \frac{G^4 M_p^5}{R_p^5\pi^2}\,,
\end{equation}
where in the above equation we took $\dot h \sim \omega_{\rm GW} h$, and the gravitational wave frequency $\omega_{\rm GW}$ was approximated by the inverse of the free-fall time $t_{\rm coll} = (3\pi/(16 G\rho(a_{\rm NL})))^{1/2}$, with $\rho(a)$ the background density of the universe (Eq.~\eqref{eq:background}). Then, the radiated GW energy is $E_{\rm GW} = L_{\rm GW}^{\rm coll }t_{\rm coll}$.

Now, to relate the above with the different examples here studied, we assume that a perturbation with a wavenumber $k$ should form a primordial structure with a mass $M_p \simeq M_{\rm HC}(k)$ (see Eq.~\eqref{mass}) and a radius $R_p \simeq [3M_p/(4\pi\rho(a_{\rm NL}))]^{1/3}$ once it becomes nonlinear. We assume this happens when $\delta(a_{\rm NL}) = \delta_{\rm HC}(k)(a_{\rm NL}/a_{\rm HC}) = 1.39$, where, as mentioned earlier, $\delta_{\rm HC}(k)$ is the amplitude of the density contrast with wavenumber $k$ evaluated at horizon crossing time and is roughly given by $\delta_{\rm HC}(k)\simeq (2/5)\mathcal{P}_\mathcal{R}(k)^{1/2}$. Taking into account that the energy density in GWs scale as $a^{-4}$, the energy density in GWs at the present time can then be written as 
\begin{equation}
    \frac{d\rho_{\rm GW}^{\rm coll,0}}{dln k } = \frac{3\rho (a_{\rm NL})}{M_p}L_{\rm GW}^{\rm coll}t_{\rm coll}\left(\frac{a_{\rm NL}}{a_0}\right)^4,
\end{equation}
where {quantities with an index $0$ are evaluated at the present time.} 
The above expression is only valid for estimating the energy density in GWs generated due to the collapse and formation of primordial structures in the post-inflationary universe. However, once these objects are formed, it is expected that they interact with each other (in addition to contributions from the rotation/vibration of the primordial structures), leading to the generation of more GW signals. To account for this effect, in Ref.~\citep{Eggemeier:2022gyo} the density of GWs at present was estimated as
\begin{equation}\label{eq:density}
    \frac{d\rho_{\rm GW}^{>\rm coll,0}}{d\ln k} = \varepsilon   \frac{d\rho_{\rm GW}^{\rm coll,0}}{d\ln k}\left(\frac{a_{\rm reh}}{a_{\rm NL}}\right)^p,
\end{equation}
where $a_{\rm reh}$ ($a_{\rm NL}$) is the value of the scale factor evaluated at reheating (formation of the primordial structure) time, and $\varepsilon = 0.4$ and $p= 1.84$ are two free parameters that were obtained numerically once matching the above formula with numerical simulations. Using the above expression, we can then calculate the current energy density parameter of GWs as
\begin{equation}\label{eq:om_gw}
    \Omega_{\rm GW,0}(k) = \frac{1}{\rho_{c,0}}\frac{d\rho_{\rm GW}^{>\rm coll,0}}{d\ln k},
\end{equation}

\begin{figure}
    \centering
    \includegraphics[width=3.5in]{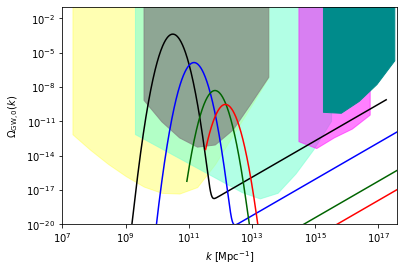}
    \caption{\footnotesize{Similar to Fig.~\ref{fig:SGWB_SO}}, but with solid lines showing the GW spectrum produced by the nonlinear structures formed during reheating.}
    \label{fig:Om_md}
\end{figure}

Following the colour codes of previous plots, in Fig. \ref{fig:Om_md}, we have plotted the SGWB produced from the PPS in Eq.~\eqref{eq:pps2} but once interchanging the Heaviside step function in that equation by  $\Theta(k-k_{\rm max})$, i.e.~we are considering only scales that reach the nonlinear regime.

\section{Prospects of detection with space- and Earth-based experiments}
\label{Sec:VI}

As discussed in the previous sections, the amplitude of the GWs generated for each scenario of PBH formation is directly related to the amplitude of the PPS. In order for PBHs to constitute all of the DM in the universe, it is necessary for the amplitude to be $\mathcal{P}_\mathcal{R}\sim 10^{-2}$ in all cases (for each reheating temperature). We have seen in the proceedings that this implies that also GWs with sufficiently large amplitude to be detected by future space- and earth-based experiments will be generated.


Indeed, although our more conservative results for perturbative second-order GWs with $\beta = 1$ indicate a SGWB that is undetectable by existing or proposed GW experiments, this is no longer the case for scales that reach the nonlinear regime, and we should expect a significant GW signal due to the different nonlinear effects that play a role during early structure formation, as can be seen in Fig.~\ref{fig:Om_md} and from Eq.~\eqref{eq:density}. For these nonlinear scales, it is even possible that we could also have reheating temperatures where the emission of GWs is so significant that it could even have a contribution comparable to the current value of relativistic species in the universe. Of course, this does not represent a plausible scenario and, therefore, that region of the model's parameter space should be ruled out. 

Additionally, if PBHs form during slow-reheating, we would also expect a shift of the peak {(or both peaks in the case of having a not to slow-reheating transition to RD)} of the SGWB spectrum to smaller frequencies {(or smaller $k$s)} than in standard radiation for the GWs generated. {This means, {for decreasing values of the reheating temperature,} and if we assume PBHs to form during slow-reheating, we expect the SGWB spectrum to peak at increasingly smaller frequencies}. However, in all cases, we found that experiments like $\mu$Ares, BBO, and LISA should be necessary in order to detect the SGWB produced by this scenario (as it is for the RD scenario). 

Note, that for certain reheating temperatures (temperatures close to the temperature of BBN, $T_{\rm reh}\rm \sim 10^{-2}~\rm{GeV}$) experiments such as BBO and ET should be capable of detecting an SGWB, even in the absence of a feature in the PPS, as was reported in Ref.~\citep{Eggemeier:2022gyo}. This is seen in the upward tails of the black curve in Fig.~\ref{fig:Om_md}.

We emphasize that we focused on the particular case where the entirety of the DM in the universe is composed of PBHs with a mass  $M_{\rm PBH}\simeq 10^{21}~\rm{g}$, leading to the various amplitudes of the SGWB presented here. However, we have seen that the SGWB spectrum scales like $\Omega_{\rm GW}\propto \mathcal{P}_\mathcal{R}^2$ for the second-order contribution for GWs and $\Omega_{\rm GW}\propto \mathcal{P}_\mathcal{R}^{5/4}$ for the nonlinear part of the spectrum. While, on the other hand, the abundances of PBHs are very sensitive to the PPS such that the PBH abundance is exponentially suppressed when dealing with a PPS with an amplitude slightly smaller than the one studied in this work. This implies that for PBHs to constitute a significant portion of DM in this mass range in other models, the amplitude of the spectrum can not be much smaller than the value studied here, and therefore the SGWB generated should not vary much from what we obtained in  the present work. In other words, if we want PBHs with a mass of $M_{\rm PBH}\simeq 10^{21}~\rm{g}$ to form a significant portion of the DM in the universe, this will lead to detectable SGWBs. Also, the position of where the peak (or the peaks) of the SGWB is detected will be intimately related to the nature of the PBH formation scenario (whether it formed during the slow-reheating scenario or not). {Specifically, in the case that PBHs formed during the slow-reheating process, we expect that the nonlinear dynamics of primordial structures (their formation and interaction) will produce an SGWB where the peak coincides with the position of the peak in the PPS that generated the PBHs. Additionally, if the reheating of the universe occurred not so slowly (the gravitational potential did not decay too much during the MD to RD transition), we would expect the generation of a second peak that approximately coincides with the scale at which universe reheating occurs.}

\section{Discussion and conclusions }\label{Sec:VII}

In this article we have studied the stochastic gravitational wave background (SGWB) that can be expected in the case of primordial black holes (PBHs) with a mass of $M_{\rm PBH}\simeq 10^{21}~\rm{g}$ constituting all of the dark matter in the universe, and that form during a slow-reheating epoch. In particular, we have adopted the formation criterion due to an effective velocity dispersion, which would originate from averaging the quantum properties of the scalar field (the inflaton) responsible for generating the reheating of the universe. For this criterion, it is assumed that a perturbation that re-entered during the early reheating epoch would collapse to form a PBH whenever its density contrast evaluated at the horizon crossing time is greater than the value given in Eq.~\eqref{eq:thr}.

In this work, we have calculated the SGWB using two different approaches, where each one is applied depending on whether the scale of interest is greater or smaller than the scale $k_{\rm max}$, where $k_{\rm max}$ is the value of the maximum scale that we expect to reach the nonlinear regime during the reheating epoch. For the case $k<k_{\rm max}$, we have used the standard procedure for calculating second-order GWs in perturbation theory. We have also considered the cases of instantaneous reheating, as well as more gradual reheating, showing that in the limit of a constant decay rate for reheating, we would not expect to have a detectable SGWB signal (which would be expected in cases of a faster transition). On the other hand, for the case where $k>k_{\rm max}$, we obtained that the expected SGWB around the peak of the primordial power spectrum (necessary for PBH formation in our model) would be large enough to be detectable by various GW experiments in the future.

{We must emphasize again that our study has made several simplifications to draw the conclusions presented here, such as choosing the specific value of $M_{\text{PBH}} \approx 10^{21}$ g, the specific width of the peak in the PPS, $\sigma_p = 0.5$, or the energy scale at the end of inflation, specified once assuming the upper limit of the tensor to scalar ratio, $r = 0.064$. Therefore, let us now discuss how our conclusions are modified once these choices are relaxed. Additionally, we have included Appendix~\ref{appendixb} to complement the conclusions that will be drawn next.}

{Firstly, relaxing the value of $M_{\rm PBH}$ for the PBHs formed in the slow-reheating scenario would imply that the peak of the PPS is shifted to different positions than those presented in this work (for the same reheating temperature), obtaining, for example, that as the mass of the formed PBHs becomes smaller (or larger), the peak would move to a higher (or lower) value of $k$. Relaxing this choice of $M_{\text{PBH}}$ to other values would also allow us to accept higher or lower values of the reheating temperature, depending on whether PBHs with smaller or larger masses are formed, respectively. This is because the only condition necessary for PBHs is that the scales that could form PBHs continue to have sufficient time to re-enter the cosmological horizon and gravitationally collapse to form PBHs. In terms of the SGWB spectrum, we should expect that the second peak, the one that should form due to the nonlinear dynamics of the inflaton, would be modified, following the shift of the PPS when assuming a different PBH mass.}

{It is important to mention that in the slow-reheating scenario, the peak shifts less as a function of the PBH mass than in the standard RD case. This can be clearly seen since, in general, we have $M_{\rm PBH} \sim H_{\rm HC}^{-1}$, thus giving us $M_{\rm PBH} \sim k^{-3}$ in the slow-reheating scenario and $M_{\rm PBH} \sim k^{-2}$ in the standard RD epoch. As a consequence of this, we would thus expect experiments like LISA to be able to test a wider range of PBH masses in the slow-reheating than in the RD case.}

{On the other hand, relaxing the adopted value of $\sigma_p$ would imply that the SGWB spectrum would also be stretched or elongated, depending on the value of $\sigma_p$. More interestingly, for a sufficiently large $\sigma_p$, we could have scenarios where the peak of the PPS reaches the nonlinear regime and thus forms PBHs, while a significant portion of the amplified spectrum remains in the perturbative regime. This, together with the Poltergeist mechanism studied in the instant reheating scenario, could lead to a detectable contribution of SGWB from the perturbative part of the spectrum by future experiments, even if the transition from MD to RD is not instantaneous (similar to what we obtained in our example with $T_{\text{reh}} = 10^3~\text{GeV}$; see Fig.~\ref{fig:SGWB_SO}).}

{Finally, the choice we made for the value of the tensor-to-scalar ratio, $r$, is not very important for the results presented in this work, since the energy scale of inflation only plays a role in imposing bounds on how long reheating must last for a scale to form a PBH, as well as giving the mass of the minimum scale that can form a primordial structure. Considering a smaller value for $r$, as expected to be in the sensitivity range of future experiments, would imply that a shorter duration of reheating is required than in this scenario of high-energy inflation. This is, again, because the only requirement needed for a scale to form a PBH during slow-reheating is that reheating lasts long enough for the scale to re-enter the cosmological horizon and grow sufficiently to form the PBH.
}

{To conclude, we have to mention that the} study detailed in this work is in large part based on different semi-analytical approximations, so numerical validation is still needed for the various conclusions drawn. It should also be noted that our work does not account for the complete history of SGWB that could be expected in an early slow-reheating scenario, as we have not included effects such as the evaporation of the CDM-like structures or the SGWB expected due to the formation, interaction, and evaporation of the soliton-like structures expected to form at the center of the CDM-like structures. We will return to these issues in future work.



\section*{Acknowledgements}
The authors are grateful to Raphael Picard for useful discussions and comments. 
 LEP and JCH acknowledge sponsorship from CONAHCyT Network Project 304001 “Estudio de campos escalares con aplicaciones en cosmología y astrofísica”. LEP is supported by a Royal Society funded postdoctoral position.  JCH acknowledges financial support from PAPIIT-UNAM programme Grant IG102123 “Laboratorio de Modelos y Datos (LAMOD) para
proyectos de Investigación Científica: Censos Astrofísicos". KAM is supported in
part by STFC grants ST/T000341/1 and ST/X000931/1. DJM is supported by a Royal Society University Research Fellowship.\\

\appendix

\section{Interpolation between a gradual and an instant transition}\label{ap:1b}

In this appendix we test whether our approximation can  describe the interpolation between a gradual and an instant transition from slow-reheating to RD. For this, we calculate the SGWB expected in the case of having an slow-roll PPS,
\begin{equation}
 \mathcal{P}_\mathcal{R}(k) = A_s\left(\frac{k}{k_*}\right)^{n_s-1}\,,
\end{equation}
where we take the preferred values for $A_s$ and $n_s$ given by Planck constraints \citep{Planck:2018jri}, and compare our results with the ones obtained in Ref.~\citep{Pearce:2023kxp}, where the interpolation between fast and gradual transition was done assuming a time-dependent decay rate of the form
\begin{equation}
    \Gamma(\eta) = \bar \Gamma\frac{\tanh(\gamma(\eta-\eta_\Gamma)+1)}{2},
\end{equation}
where the case $\gamma = 0$ coincides with the limit of a constant decay rate whereas the limit $\gamma\rightarrow \infty$ is the one for an instant transition. To have comparable results to the ones reported in Ref.~\citep{Pearce:2023kxp}, we take  $\eta_{\rm reh} = 450/k_{\rm max}$ and we used the formalism described in Sec.~\ref{sec:second_o} to calculate the SGWB in instant and more gradual transitions. 

\begin{figure}
    \centering
    \includegraphics[width=3.5in]{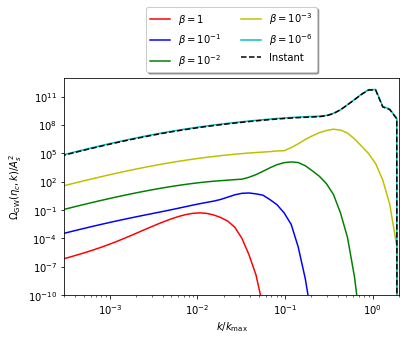}
    \caption{\footnotesize{SGWB generated by an slow-roll PPS for different values of the $\beta$ parameter (see Sec.~\ref{sec:second_o}.)}}
    \label{fig:inst_grad}
\end{figure}

Our results are depicted in Fig.~\ref{fig:inst_grad}, where we have explored various values of the $\beta$ parameter in Eq.~\eqref{eq:beta_par}. Upon comparing our plot with Fig.~3 in Ref.~\citep{Pearce:2023kxp}, we observe that our approximation effectively captures the key features of the SGWB illustrated in their figure. Specifically, as the $\beta$ parameter increases (approaching a scenario of constant decay rate), we observe a sharp decline in the SGWB for longer wavelengths. Furthermore, it is evident that with larger values of $\beta$, this decay occurs at smaller $k$ values.

Comparing the figures further, it is important to mention that we found two main differences, which we will discuss below. Firstly, in Fig.~3 in Ref.~\citep{Pearce:2023kxp}, it can be observed that the abrupt decay of the SGWB spectrum ceases at some point, continuing to decay but at a slower rate. This behavior is not observed in our figure, so we attribute that part of the spectrum precisely to the term $I_{\rm reh}$, which we ignored when calculating the SGWB. However, since this latter part of the spectrum would correspond to amplitudes well below the sensitivity regions of experiments, we consider that we are not losing any significant contribution regarding how we make our estimates.

Secondly, we can observe that our case with $\beta = 1$ has a slightly lower maximum amplitude in the SGWB compared to what the authors reported in Ref.~\citep{Pearce:2023kxp} (in both cases the maximum of the SGWB are $\sim O(10^{-1}$)), and at lower wave numbers $k$, the spectrum decays less rapidly than in our case. Again, we attribute this discrepancy to the term $I_{\rm reh}$ that we did not consider for our calculation (for behavior at small values of $k$), as well as to the fact that we are assuming the decoupling time of the gravitational potential from matter perturbations occur exactly when $|\ddot \Phi/\Phi|_{t = t_d}\simeq k^2/(3a(\eta_d)^2)$, which is not necessarily true and should instead only give us a minimum bound for the expected SGWB for each case.

However, since the maxima for the case of a constant decay rate are of the same order in both figures and considering that the different cases of more gradual decay were able to correctly describe the important features (e.g.~the shape of the gravitational wave spectra), we consider that our approach to calculating the SGWB during a gradual reheating is sufficiently good to draw conclusions regarding the different examples studied in this work. Furthermore, our more phenomenological approach has the significant advantage of not assuming a functional form for the decay rate of the inflaton, instead aiming to draw general conclusions solely based on the amplitude of the gravitational potential at the time of its decoupling from matter perturbations.

\section{Varying the parameters $\sigma_p$ and $M_{\rm PBH}$}\label{appendixb}

\begin{figure}
    \centering
    \includegraphics[width=3.6in]{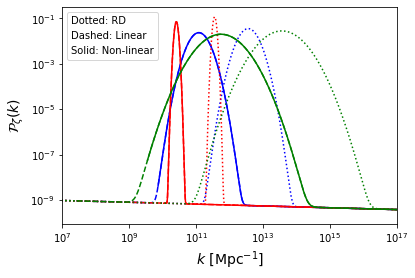}\\
    \includegraphics[width=3.6in]{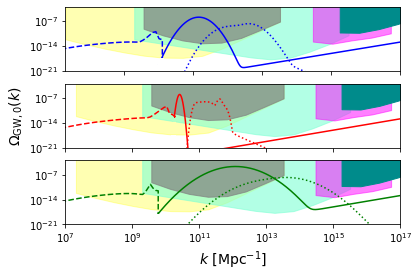}
    \caption{\footnotesize{Upper plot: We show the PPS for three different examples with $M_{\rm PBH} = 10^{21}~\rm{g}$ and $\sigma_p = 0.5$ (in blue), $M_{\rm PBH} = 10^{23}~\rm{g}$ and $\sigma_p = 0.1$ (in red), and $M_{\rm PBH} = 10^{19}~\rm{g}$ and $\sigma_p = 1$ (in green). The solid and dashed lines correspond to the scales that reach the nonlinear regime and the ones that reenter the cosmological horizon but stay in the perturbative regime, respectively. We have also included the corresponding PPS (in dotted lines) for the same parameters of $M_{\rm PBH}$ and $\sigma_p$ but this time by assuming the PBHs formed during the standard RD scenario. Lower plot: We show the SGWB associated with each of the PPSs. For the perturbative regime of the slow-reheating, we are only showing the most promising scenario of an instant transition. Notice that we have also included in dotted lines the SGWB that we should expect in the case in which the PBHs formed during standard RD.}}
    \label{fig:appendix2}
\end{figure}

All our main conclusions drawn in this paper have been given once requiring the mass of all of the DM in the universe is made by PBHs formed during slow-reheating with a mass of $M_{\rm PBH}\simeq 10^{21}~\rm{g}$ and by fixing the parameter $\sigma_p = 0.5$. In this appendix, we are interested in showing how the spectra of SGWBs change once we consider a different set of parameters. In particular, in the upper plot of Fig.~\ref{fig:appendix2} we show three different PPSs in which we have adopted the values $M_{\rm PBH} = 10^{21}~\rm{g}$ and $\sigma_p = 0.5$ (in blue), $M_{\rm PBH} = 10^{23}~\rm{g}$ and $\sigma_p = 0.1$ (in red), and $M_{\rm PBH} = 10^{19}~\rm{g}$ and $\sigma_p = 1$ (in green). For all the cases, we have considered that the reheating of the universe took place at $T_{\rm reh} = 1~\rm{GeV}$ or the case of an instant reheating scenario (dotted curves). Then, at the bottom of the Figure we have also calculated the SGWB expected in each of the models, where for the perturbative regime of the SGWB for the slow-reheating, we are showing only the most promising scenario of an instant transition from early reheating to RD (see Sec.~\ref{Sec:V}). 

\bibliographystyle{ieeetr}
\bibliography{biblio}
\end{document}